\documentstyle{article}

\newcommand{\CPN}{{\rm CP}^{N-1}}
\newcommand{\Tr}{\mathop{\rm Tr}}

\newcommand\simle{\rlap{\raise 2pt \hbox{$<$}}{\lower 2pt \hbox{$\sim$}}}

\title{\rightline{\tenrm IFUP-TH 46/92}\bigskip
Scaling and asymptotic scaling in two-dimensional $\CPN$ models
\footnote{Talk presented at the {\em Lattice '92\/} Conference, Amsterdam.}
}

\author{%
Massimo Campostrini, Paolo Rossi, and Ettore Vicari\\
\tenit Istituto Nazionale di Fisica Nucleare, Sezione di Pisa,\\
\tenit Dipartimento di Fisica dell'Universit\`a, I-56126 Pisa, Italy.%
}

\date{October 6th, 1992}

\begin{document}

\maketitle

\begin{abstract}
Two-dimensional $\CPN$ models are investigated by Monte Carlo methods on
the lattice, for values of $N$ ranging from 2 to 21.  Scaling and rotation
invariance are studied by comparing different definitions of correlation
length $\xi$. Several lattice formulations are compared and shown to enjoy
scaling for $\xi$ as small as $2.5$.

Asymptotic scaling is investigated using as bare coupling constant both the
usual $\beta$ and $\beta_E$ (related to the internal energy); the latter is
shown to improve asymptotic scaling properties.
Studies of finite size effects show their $N$-dependence to be highly
non-trivial, due to the increasing radius of the $\bar z z$ bound states at
large $N$.
\end{abstract}

\section{Introduction}
Two-dimensional $\CPN$ models share many dynamical features with
four-dimensional ${\rm SU}(N)$ gauge theories, but are much easier to
explore numerically and are amenable to analytical analysis by means of the
$1/N$ expansion.  $\CPN$ models can therefore be useful as a test-bed for
algorithms and numerical methods to be applied to lattice QCD, since
systematic errors can be explored in details and numerical results can be
compared with precise theoretical predictions. Indeed several studies of
$\CPN$ and related models have been presented in this Conference.

In this paper we summarize studies of the scaling, asymptotic scaling, and
finite-size scaling properties of different lattice formulations of $\CPN$
models. The full details of the simulations and analyses of different
aspects of the models can be found in Refs.~\cite{cpn1,cpn2}.

Many recent lattice studies of $\CPN$ models used the ``quartic'' action
\begin{equation}
S_1 = -N\beta\sum_{n,\mu}\left|\bar z_{n+\mu}z_n\right|^2.
\end{equation}
We performed most of our simulations using the ``gauge'' action
\begin{equation}
S_{\rm g} = -N\beta\sum_{n,\mu}\left(
   \bar z_{n+\mu}\lambda_{n,\mu}z_n +
   \bar z_n\bar\lambda_{n,\mu}z_{n+\mu}\right).
\end{equation}
$z_n$ is an $N$-component complex scalar field constrained by the
condition $\bar z_nz_n = 1$, and $\lambda_{n,\mu}$ is a ${\rm U}(1)$ gauge
field. $S_{\rm g}$ enjoys several advantages over $S_1$, the most important
being a better scaling behavior. We also considered the tree-level
Symanzik-improved versions of both actions, which will be denoted by
$S_1^{\rm Sym}$ and $S_{\rm g}^{\rm Sym}$.

We performed Monte Carlo simulations of the ${\rm CP}^1$, ${\rm CP}^3$,
${\rm CP}^9$, and ${\rm CP}^{20}$ models, for values of the correlation
length $\xi$ ranging from 2.5 to 30.

\section{Correlation Length and Scaling}
The most interesting correlation function of the model selects the
${\rm SU}(N)$-adjoint, gauge-neutral channel:
\begin{eqnarray}
\lefteqn{G_P(x) = \left<\Tr\{P(x) P(0)\}\right>_{\rm conn}\,,} \nonumber \\
\lefteqn{P_{ij}(x) = \bar z_i(x) z_j(x).}
\end{eqnarray}
We defined the usual wall-wall correlation length $\xi_{\rm w}$ from the
exponential decay of the zero-momentum projection of
$G_P$. In order to check for rotation invariance, we also defined the
diagonal wall-wall correlation length $\xi_{\rm dw}$ in the obvious way.
Both $\xi_{\rm w}$ and $\xi_{\rm dw}$ tend to the inverse mass gap in
the scaling limit.  The ratio $\xi_{\rm dw}/\xi_{\rm w}$ is plotted in
Figs.~1a, 2a, 3, and 4 for different values of $N$ and different actions,
showing rotation invariance even for the smallest values of $\xi$.

We also chose a different definition of correlation length:
\begin{equation}
\xi_G^2 = {1\over4\sin^2\pi/L} \,
\left[{\widetilde G_P(0,0)\over\widetilde G_P(0,1)} - 1\right],
\end{equation}
where $\widetilde G_P(p_1,p_2)$ is the Fourier transform of $G_P$;
$\widetilde G_P(0,0)$ is the magnetic susceptibility $\chi$.
In the scaling limit, $\xi_G$ tends to the second moment of the correlation
function. The physical quantity $\xi_G/\xi_{\rm w}$ must be independent of
$\beta$ and on the choice of the action in the scaling region.
$\xi_G/\xi_{\rm w}$ is plotted in Figs.~1b, 2b, 3, and 4 for different
values of $N$ and different actions, showing scaling for all the values of
$\xi$. In the following we will choose $\xi_G$ as length scale, since it
is less noisy then $\xi_{\rm w}$ or $\xi_{\rm dw}$.

\section{Asymptotic Scaling}
We compared the $\beta$-dependence of the lattice mass scale $1/\xi_G$ with
the two-loop perturbative formula $f(\beta) =
(2\pi\beta)^{2/N}\exp(-2\pi\beta)$, by considering the ratio
$M_G/\Lambda_{\rm latt}\equiv[\xi_Gf(\beta)]^{-1}$. In the asymptotic
regime, $M_G/\Lambda_{\rm latt}$ must be constant plus corrections
$O(1/\beta)$; the $\beta\to\infty$ limit may therefore be approached very
slowly. In order to compare different lattice formulations, we reexpressed
the specific $\Lambda_{\rm latt}$ of each formulation in terms
$\Lambda_{\rm g}$.

We used two different schemes for the analysis: in the ``standard'' scheme
the inverse bare coupling is simply $\beta$, the action parameter. In the
``energy'' scheme, suggested by G.~Parisi \cite{Parisi}, the inverse bare
coupling is $\beta_E\equiv 1/(2E)$; $E$ is the expectation value of the
link action, normalized to $E(\beta\,{=}\,\infty)=0$. At the perturbative
level, $\beta$ and $\beta_E$ are simply two different definitions of
coupling constant. We reexpressed $\Lambda_{\rm latt}$ in the ``energy''
scheme in terms of $\Lambda_{\rm g}$ in the ``standard'' scheme.

 The results for ${\rm CP}^1$, ${\rm CP}^3$, and ${\rm CP}^9$, are
presented in Figs.~5--7.  Results for the ``standard'' scheme show
asymptotic scaling violations and disagreement between different lattice
actions, but both effects are compatible with the expected $1/\beta$
behavior. On the other hand, results for the ``energy'' scheme are
remarkably $\beta$- and action-independent.  In the case of ${\rm CP}^1$,
the ``energy'' scheme result is in agreement with the exact result of
Ref.~\cite{Hasenfratz}:
\begin{equation}
{M_G\over\Lambda_{\rm g}} =
{8\over e}\,\sqrt{32}\exp\left(\pi\over4\right) \cong 36.51\,.
\end{equation}

\section{Finite Size Scaling}
We examined finite size scaling effects by studying the finite size scaling
function of an observable ${\cal O}$:
$f_{\cal O}(L/\xi) \equiv {\cal O}_L(\beta)/{\cal O}_\infty(\beta)$,
where ${\cal O}_L$ is the expectation value of the observable ${\cal O}$
measured on a $L{\times}L$ lattice.  In the scaling region, $f_{\cal O}$
must be a {\it universal\/} function of $L/\xi$, independent of $\beta$ and
of the lattice action. We present here results for the magnetic
susceptibility $\chi$; $f_\xi$ has a behavior very similar to $f_\chi$.

We expect a qualitative difference in the finite size behavior of $\CPN$
models at small and large $N$. Finite size effects for ${\rm CP}^1$ are
expected to be dominated by the inverse mass gap, therefore following the
predictions of Ref.~\cite{Luscher}:
$f_{\cal O} = 1 + O(\exp\{-L/\xi\})$.
$f_\chi$ is plotted in Fig.~8, and it is in agreement with the above
prediction.

No clear theoretical predictions are available for the ${\rm CP}^3$ model.
$f_\chi$ is plotted in Fig.~9.

The large-$N$ expansion predicts a radius of the lowest-lying state
proportional to $\xi N^{1/3}$ \cite{Witten}. Therefore we expect, for $N$
large enough, finite size effects to be dominated by the bound state radius
rather then by the inverse mass gap. Quantities like $\xi$ and $\chi$ are
expected to increase when $L$ is decreased, in contrast with the ${\rm
CP}^1$ case.  Moreover the function $f_{\cal O}\bigl(L/(\xi N^{1/3})\bigr)$
should not depend on $N$ (at least for $N\simle100$; for huge values of
$N$, the bound state decouples from $\chi$ and $\xi_G$).
Fig.~10 shows $f_\chi$ plotted as a function of $L/(\xi N^{1/3})$, for both
${\rm CP}^9$ and ${\rm CP}^{20}$, and it is in agreement with the large-$N$
expansion.

\newpage
\def\thesection{}
\section{Figure captions}
\bigskip\par\noindent
{\bf Fig.~1: }Scaling tests for ${\rm CP}^1$.
\bigskip\par\noindent
{\bf Fig.~2: }Scaling tests for ${\rm CP}^3$.
\bigskip\par\noindent
{\bf Fig.~3: }Scaling tests for ${\rm CP}^9$.
\bigskip\par\noindent
{\bf Fig.~4: }Scaling tests for ${\rm CP}^{20}$.
\bigskip\par\noindent
{\bf Fig.~5: }Asymptotic scaling tests for ${\rm CP}^1$.
\bigskip\par\noindent
{\bf Fig.~6: }Asymptotic scaling tests for ${\rm CP}^3$.
\bigskip\par\noindent
{\bf Fig.~7: }Asymptotic scaling tests for ${\rm CP}^9$.
\bigskip\par\noindent
{\bf Fig.~8: }Finite size scaling for ${\rm CP}^1$.
\bigskip\par\noindent
{\bf Fig.~9: }Finite size scaling for ${\rm CP}^3$.
\bigskip\par\noindent
{\bf Fig.~10: }Finite size scaling for ${\rm CP}^9$ and ${\rm CP}^{20}$.

\end{document}